# MAISY: Motion-Aware Image SYnthesis for Medical Image Motion Correction

Andrew Zhang[1], Hao Wang[1], Shuchang Ye[1], Michael Fulham[2] and Jinman Kim[1]*

[1] School of Computer Science,
University of Sydney
[2] Department of Molecular Imaging, Royal Prince Alfred Hospital
`jinman.kim@sydney.edu.au`

**Abstract.** Patient motion during medical image acquisition causes blurring, ghosting, and distorts organs, which makes image interpretation challenging. Current state-of-the-art algorithms using Generative Adversarial Network (GAN)-based methods with their ability to learn the mappings between corrupted images and their ground truth via Structural Similarity Index Measure (SSIM) loss effectively generate motion-free images. However, we identified the following limitations: (i) they mainly focus on global structural characteristics and therefore overlook localized features that often carry critical pathological information, and (ii) the SSIM loss function struggles to handle images with varying pixel intensities, luminance factors, and variance. In this study, we propose Motion-Aware Image SYnthesis (MAISY) which initially characterize motion and then uses it for correction by: (a) leveraging the foundation model Segment Anything Model (SAM), to dynamically learn spatial patterns along anatomical boundaries where motion artifacts are most pronounced and, (b) introducing the Variance-Selective SSIM (VS-SSIM) loss which adaptively emphasizes spatial regions with high pixel variance to preserve essential anatomical details during artifact correction. Experiments on chest and head CT datasets demonstrate that our model outperformed the state-of-the-art counterparts, with Peak Signal-to-Noise Ratio (PSNR) increasing by 40%, SSIM by 10%, and Dice by 16%.

## 1 Introduction

Patient motions encompasses physical movements that can take place during the acquisition of medical imaging scan [1]. Motion consists of two categories: (i) voluntary motion: which can arise from the contraction of muscles such as the head, neck, or limb and, (ii) involuntary motion: which can arise from contractions of muscles such as the heart and movement of the bowl due to peristalsis. These movements can result in motion artifacts in imaging, creating unwanted blurring, ghosting, and distortions of anatomical and functional regions. There has been sustained research effort in characterizing and correcting for motion artifacts. Examples include prospective techniques during the acquisition of scans, such as the gating of organs and optical tracking of patient motion [2]. In contrast, retrospective methods are 'image-based', which corrects for motion artifacts after image acquisition.

Deep learning algorithms are the state-of-the-art in image-based motion correction, with generative networks – methods that generate motion-free images based on the mappings between corrupted images and their ground truths – leading the performance. Early generative methods relied on an autoencoder to reduce motion artifacts which worked by reconstructing a motion-free 'synthetic' image through learning motion-induced distortions, e.g., for brain magnetic resonance (MR) images [3]. While autoencoders learn to remove motion artifacts, Generative Adversarial Network (GAN)-based methods improved the learning process with an additional network to progressively refine synthetic image generations using ground truths reference to generate motion-free images. This process captured complex anatomical details by learning non-linear motion distortions while retaining anatomical/functional structures [4]. Conditional GAN (cGAN) and Cycle GAN were further introduced to enhance GAN. cGAN-based methods often use a U-Net architecture as the generator and a patch GAN discriminator [5-8]. Here, U-Net was responsible for mapping motion-corrupted CT images into motion-free target domains, while the patch GAN discriminator worked together with the U-Net to generate a pixel-realistic image. On the other hand, Cycle GAN-based methods often use various generator architectures, including convolutional networks with self-attention mechanisms [9], autoencoders [10], and U-Net architectures [11], with patch GAN as the discriminator. These methods relied on bidirectional mappings between motion image and non-motion target domains, allowing the U-Net generator to remove motion artifacts in images and patch GAN to generate realistic outputs.

However, existing methods focused on global structural consistency while neglecting local motion-specific patterns [5, 6, 9]. These artifacts obscured pathological features which are diagnostically important. Additionally, generative algorithms commonly used the Structural Similarity Index Measure (SSIM) loss, which was found to be inaccurate for image-to-image comparisons made in GANs due to its inability to address images with varying pixel intensities, luminance factors, and variance [12]. To address these limitations, we propose Motion-Aware SAM-based Image SYnthesis (MAISY) with the following contributions via two modules:

1. MAISY leverages the Segment Anything Model (SAM), a large-scale pre-trained foundation model [13], to extract mask-based feature representations that capture spatial patterns around anatomical boundaries where motion artifacts predominantly occur. This is the first work to exploit the powerful pre-trained foundation model adapted for solving patient motion correction.
2. We introduce a novel Variance-Selective SSIM (VS-SSIM) loss function that decomposes the image into localized patches and selectively emphasizes regions with high pixel variance, indicative of anatomical regions vulnerable to severe motion artifacts. By applying SSIM at the patch level, our proposed loss mitigates the distortions caused by intensity and luminance variations.

Experiments on head and chest CT XCAT phantom datasets show that our MAISY achieves state-of-the-art performance. We demonstrate the importance of our modules with an ablation study and a modularity study where we show the benefit of adding our modules to other motion correction methods. Furthermore, MAISY performed consistently on both datasets whilst comparison methods were only trained and tested on one anatomical region with low performance when fine-tuned on another anatomical domain [5-7, 9, 10].

## 2    Methodology

An overview of MAISY is presented in Fig. 1 using a chest CT dataset: SAM-based image segmentation is used to produce a mask feature maps which is concatenated with the CT images and used as input into an adapted Attention Guided GAN (AGGAN) obtained from Tang, et al. [14] with the VS-SSIM loss function. The output is a motion corrected CT image. We generate motion corrected image, $\hat{y}$, according to:

$$\hat{y} = \arg\min[\lambda_a L_{VS-SSIM}(y_{generated}, y_{gt}) + \lambda_b \beta(y_{generated})]. \quad (1)$$

The function $L_{VS-SSIM}(y_{generated}, y_{gt})$ is the VS-SSIM within the AGGAN which compares patches between the generated image $y_{generated}$ and the ground truth image. $\beta(y_{generated})$ is the additional regularizations such as loss functions. The λ represents the balancing term to provide importance to each constraint.

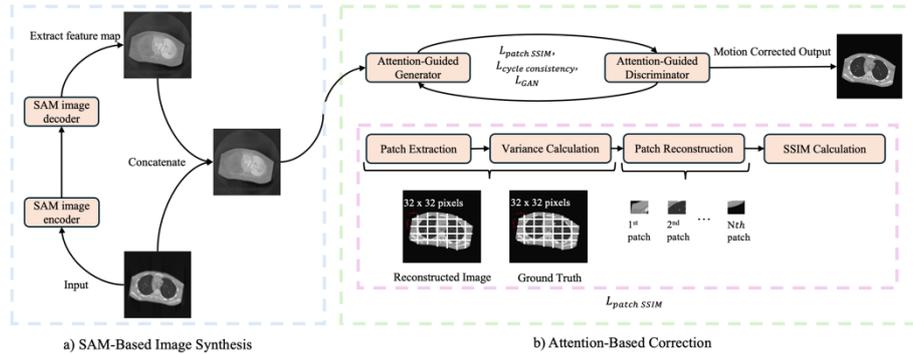

**Fig. 1.** Overall framework of MAISY

### 2.1    SAM Based Image Synthesis

The first approach to MAISY is the use of SAM to segment the anatomical regions. Here, the image encoder extracts the image embeddings from the image which is then passed through a mask decoder. We provided point prompts within the CT image where the image embedding is passed into a mask decoder where a pre-binarized mask feature map is extracted from the image. The feature map guides the AGGAN to focus on correcting within the anatomical boundary by supplying sematic information on the segmentation mask surrounding the anatomical region. Assuming $y_{corrupted}$ is the input into SAM:

$$F = D(E(y_{corrupted}), P) \qquad (2)$$

The mask feature map *F* is extracted from the SAM image decoder *D*, which has processed the encoded input $E(y_{corrupted})$ based on the manual user prompts *P*.

### 2.2  Attention-Based Motion Correction

The second step to MAISY is where we concatenate the input image $y_{corrupted}$ with the mask feature map *F* (from Section 2.1) through the channel dimension and then use it as the input into an AGGAN [14].

$$C = Concat(y_{corrupted}, F) \qquad (3)$$

Here, the concatenated image, *C*, is passed into the AGGAN generators where deep residual blocks with skip connections, content mask, $R_{generated}$ and attention mask $M_{generated}$ highlight regions perceived to be important. Based on this information, the generator outputs a motion-corrected CT image $y_{generated}$ to the discriminators.

$$y_{generated} = G(R_{generated} * M_{generated} + C * (1 - M_{generated})) \qquad (4)$$

These were modified for image-to-image translation tasks, with a patch GAN and pixel discriminator working together to capture global and local features, comparing the generated output against an unpaired motion-free target image. Unlike the original AGGAN, our model implemented an additional motion-specific loss, the VS-SSIM loss function.

$$VS\text{-}SSIM(y_{gt}, y_{generated}) = \frac{1}{|\Omega|} \sum_{i \in \Omega} SSIM(y_{gt}^i, y_{generated}^i) \qquad (5)$$

$$L_{VS-SSIM}(y_{generated}, y_{gt}) = 1 - VS\text{-}SSIM(y_{gt}, y_{generated}) \qquad (6)$$

The VS-SSIM loss function ranks the pixel variance of all patches where $\Omega$ represents the top patches with the highest pixel variance. SSIM loss is calculated across the $\Omega$ patches and averaged to calculate a high variance patch structural loss. We use $1 - VS\text{-}SSIM(y_{gt}, y_{generated})$ to ensure minimization when calculating loss function. The overall loss function combines the VS-SSIM loss $L_{VS-SSIM}$ with GAN loss $L_{GAN}$ and cycle loss $L_{cycle}$.

$$Total\ Loss = \lambda_a L_{VS-SSIM} + \lambda_b (L_{GAN} + L_{cycle}) \qquad (7)$$

Most importantly, the combination of both modules increases MAISY's ability to focus on motion-related anatomical areas. As illustrated in Fig. 1b, the SAM feature map can complement the patching process in the loss function, VS-SSIM, allowing patching only within the segmented chest region.

## 3    Dataset and Experiment

### 3.1    Data Preprocessing

We used motion simulated XCAT phantom of the CT chest and combined head and jaw datasets [15]. Motion artifacts were simulated with step-and-shoot flat-bed CT (FBCT) geometry with six degrees of freedom [15]. For the chest dataset, rigid motions were simulated with the cubic-beta-spline interpolation method, and non-rigid motions of simulated chest expansion during breathing. 2000 scans were derived from 14 patients. For the head and jaw dataset, rigid head motions were also simulated using the cubic-beta-spline interpolation method. The raw projection data were reconstructed using filtered backpropagation, resulting in 256×256 image size with pixel dimension of 1mm×1mm. We normalized the pixel values of the concatenated image and mask feature map to a range of -1 and 1. The dataset was split into three sections for both datasets. For the chest dataset, we used 1650 scans for training, 250 for validation, and 100 for testing. For the head and jaw dataset, we used 20 scans for further training, 5 for validation and 20 for testing.

### 3.2    Experiment Design

To evaluate performance against state-of-the-art algorithms, a comparison study was made between U-Net and cGAN adapted from Usui, et al. [5], Pix2Pix model adapted from Isola, et al. [16] and AGGAN. We adopted established image comparison metrics in motion correction studies: PSNR, SSIM and Dice Score [5, 6, 8-11]. PSNR measures pixel fidelity, SSIM score structural similarity and Dice Score measures segmentation performance. Dice was calculated by thresholding the segmentation result to create a binary mask prior to comparison to the ground truth mask of the overall structure. An ablation study was conducted by assessing the effects of different modules and their effects on the evaluation metrics. Further, a modularity study was done where SAM Masking and VS-SSIM modules were added to different image-to-image translation models of Pix2Pix and cGAN.

### 3.3    Implementation Details

The MAISY was implemented using Python using Nvidia GeForce GTX 3090 Ti with 24 GB VRAM. For each epoch, five images were used for training and one randomly selected image for validation. All models used on the chest dataset were trained to 100 epochs with a batch size of 5 and a learning rate of 0.0002. Further, 10 epochs of fine-tuning was implemented on the head and jaw dataset.

## 4    Results and Discussion

### 4.1    Comparison with the state-of-the-art

Table 1 presents the quantitative comparison results between MAISY and the state-of-the-art methods. MAISY achieved the highest SSIM (0.989) and PSNR (43.559) scores in the chest dataset, suggesting that the majority of the motion was corrected while maintaining spatial and structural similarity. The margin to the second-best performing method, AGGAN, was substantial, with a 3.20% increase in PSNR and a 0.71% increase in SSIM for the chest dataset and a 13.91% increase in PSNR and a 1.34% increase in SSIM for the head and jaw dataset. Additionally, MAISY also obtained a Dice of 0.991, a 0.81% increase for the chest dataset and a Dice of 0.999, a 0.10% increase for the head and jaw dataset, which indicates a strong overlap between our MAISY-generated image and the ground truth.

Table 1. Comparison of state-of-the-art motion correction methods and MAISY for chest and head & jaw CT datasets. Best results are in **bold** and second-best underlined.

| Methods | Chest | | | Head & Jaw | | |
|---|---|---|---|---|---|---|
| | PSNR | SSIM | Dice | PSNR | SSIM | Dice |
| U-Net | 20.156 | 0.367 | 0.667 | 19.761 | 0.400 | 0.577 |
| cGAN | 31.210 | 0.901 | 0.855 | 25.482 | 0.701 | 0.637 |
| Pix2Pix | 21.344 | 0.712 | 0.349 | 22.309 | 0.740 | 0.409 |
| Cycle GAN | 20.968 | 0.656 | 0.458 | 25.720 | 0.668 | 0.596 |
| AGGAN | <u>42.211</u> | 0.982 | <u>0.983</u> | <u>38.075</u> | 0.967 | <u>0.998</u> |
| MAISY | **43.573** | **0.989** | **0.991** | **43.372** | **0.980** | **0.999** |

We attribute the improvements of MAISY to its motion localization strategy. The synergy between SAM and VS-SSIM enables MAISY to avoid global smoothing during correction and focus on motion artifacts surrounding the anatomical regions defined by the segmentation mask. Additionally, the semantic segmentation by SAM provides a pre-attention guide that directs the attention mask in AGGAN to build upon it through constant refinements during adversarial training. This allows MAISY to focus on areas where correction is needed, preventing the degradation of performance by external noises. Furthermore, the combination of a hybrid loss function with cycle consistency loss takes advantage of its ability to preserve overall image structure, achieving global and local anatomical fidelity. Despite minimal fine-tuning on only 20 head and jaw images for only 10 epochs, MAISY exhibited excellent knowledge of motion features and can easily adapt to other motion-corrupted anatomical domains.

### 4.2 Ablation Study

**Table 3.** Ablation test on the addition of modules in Attention GAN. The results of the best-performing model are in **bold**.

| Methods | PSNR | SSIM | Dice |
| --- | --- | --- | --- |
| AGGAN | 42.211 | 0.982 | 0.983 |
| AGGAN + SAM Masking | 42.126 | 0.985 | 0.990 |
| AGGAN + VS-SSIM | 43.070 | 0.987 | 0.984 |
| MAISY | **43.573** | **0.989** | **0.991** |

Table 3 summarizes the results of our ablation study, which shows that the addition of modules increased performance in all the metrics. The addition of semantic mask information by SAM aligns the mean and variance values of the ground truth image and the generated image, leading to an increase in SSIM and Dice Score. However, we see a small decrease in PSNR from 42.211 to 42.126 with SAM Masking. This is expected where the masking itself could potentially introduce a small intensity shift in AGGAN. On the other hand, the VS-SSIM acts as an adaptive regularizer which dynamically weighs patches according to their variance. This adaptation reduces local reconstruction errors during synthetic image generation on motion across all types of images, effectively increasing all metrics.

### 4.3 Modularity Study

**Table 4.** Modularity study on the effects of integrating SAM Mask Features and VS-SSIM loss function into baseline motion correction techniques

| Methods | PSNR | SSIM | Dice |
| --- | --- | --- | --- |
| cGAN | 31.210 | 0.901 | 0.855 |
| cGAN + SAM Masking | 31.598 | 0.910 | 0.863 |
| cGAN + VS-SSIM | 31.670 | 0.910 | 0.867 |
| cGAN + SAM Masking + VS-SSIM | **31.827** | **0.915** | **0.867** |
| Pix2Pix | 21.344 | 0.712 | 0.349 |
| Pix2Pix + SAM Masking | 29.601 | 0.866 | 0.813 |
| Pix2Pix + VS-SSIM | 29.733 | **0.903** | 0.811 |
| Pix2Pix + SAM Masking + VS-SSIM | **30.105** | 0.891 | **0.847** |
| Cycle GAN | 20.968 | 0.656 | 0.458 |
| Cycle GAN + SAM Masking | 21.554 | 0.765 | 0.452 |
| Cycle GAN + VS-SSIM | **21.741** | 0.751 | 0.461 |
| Cycle GAN + SAM Masking + VS-SSIM | 21.677 | **0.776** | **0.461** |

Table 4 shows performance improvement from the addition of our modules. We discovered that SAM Masking with VS-SSIM improved PSNR, SSIM, and Dice Scores in cGAN, Cycle GAN, and Pix2Pix. These architectures often focus on the global aspect of the image i.e. global losses, as such, the incorporation of patching and localization properties improves performance in generating higher-quality outputs with enhanced structural integrity by utilizing additional contextual information which had been reinforced by numerous studies [17-19].

### 4.4   Qualitative Study

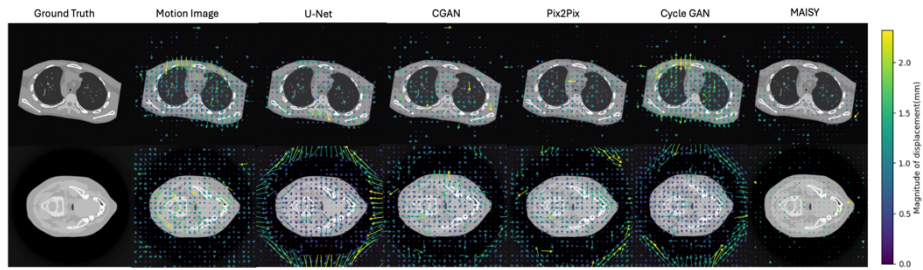

**Fig. 2.** Visualization of motion correction methods.

In Fig. 2 we show the qualitative comparisons of MAISY against the state-of-the-art correction methods with arrows delineating motion vectors. We observed that MAISY exhibited the least motion displacement vectors within the lungs and head, highlighting its performance and robustness to different anatomical domains.

## 5    Limitations

Our study was built on GAN architecture. However, GAN could be replaced with other generative image methods, such as diffusion models, to potentially further improve motion correction. We evaluated MAISY on two different body sections with varying motion artifacts, but both were CT. Our future work will investigate generalizability to other imaging modalities e.g., MR and positron emission tomography (PET) and multi-modal PET-CT.

## 6    Conclusion

We presented MAISY, a generative deep learning method, that we used to characterize and correct patient motion in medical images. We introduced attention masks in AGGAN to complement SAM for motion correction, coupled with a new VS-SSIM loss function. Our evaluations demonstrated that our method outperformed the state-of-the-art comparison methods in different anatomical domains, and that our ablation and modularity study demonstrated the importance and versatility of our proposed modules in improving image-based motion correction.